# Reference environments: A universal tool for reproducibility in computational biology

*Daniel G. Hurley[1,2,3], Joseph Cursons[1,4], Matthew Faria[1,2,3], David M. Budden[1,3], Vijay Rajagopal[1,3], Edmund J. Crampin[1,2,3,5]*

*[1]Systems Biology Laboratory, School of Mathematics and Statistics and Melbourne School of Engineering, University of Melbourne, Parkville, Victoria 3010, Australia*

*[2]ARC Centre of Excellence in Convergent Bio-Nano Science and Technology, Melbourne School of Engineering, University of Melbourne*

*[3]Department of Biomedical Engineering, Melbourne School of Engineering, University of Melbourne*

*[4]Walter and Eliza Hall Institute of Medical Research, Parkville, Melbourne*

*[5]School of Medicine, Faculty of Medicine Dentistry and Health Sciences, University of Melbourne*

daniel.hurley@unimelb.edu.au

## Introduction

Computational research should be among the most reproducible methods in the life sciences; compared to a typical laboratory method, the environment and protocol for a computational result can be specified to a much greater degree, and with much greater precision. Unfortunately, many published computational results are still difficult to reproduce[1], and methods are often not specified in sufficient detail to enable replication or reproduction of results, even those published in journals with explicit policies on making results reproducible[2]. Even the definition of reproducibility is a matter of debate; the general notion of reproducibility in experimental research is similar but not identical to that in computational research, and terms such as 'replicability', 'recomputability', 'inferential reproducibility', and 'methods reproducibility' have all been proposed to distinguish between different activities[3,4]. In this work we use the term 'computational reproducibility' to refer specifically to the challenge of recreating a published computational analysis in the same manner as it was described to have been done in the publication.

Awareness of the importance of research reproducibility has been steadily increasing in recent years, resulting in a community focus on standards for reproducibility in computational research[5–7], and on platforms[8,9] and best practices[3,10] to support it. However, making published computational research commonly and readily reproducible poses significant challenges in the areas of scholarship and research ethics[11], legal/policy frameworks[12] and technology, and it is the technological challenge we discuss here.

Many journals now require the executable code used to produce a result to be publically available as a condition of publishing. However, reproducing a specific published result requires more than the executable code used originally to produce it; it requires an entire software stack composed of all the libraries or tools upon which the code depends, and correct configuration of all components within that stack. In a modern computer system, these dependencies are constantly changing, making it very difficult to produce a precise and comprehensive specification of the method for a computational result.

Responses to this technological challenge of reproducibility in computational research are numerous and varied, and new ones are published every day; the container system Docker is popular in the bioinformatics community[13], while ad-hoc virtual machines are sometimes made available by major consortia (e.g. the ENCODE consortium virtual machine[14]), and other researchers argue for the use of 'literate programming approaches' such as iPython/Jupyter for the Python language[15], or Sweave for R[16]. Table 2 contains a more exhaustive list of the wide range of tools, platforms, and standards associated with reproducible research. Each of these approaches has merit, and any of them, if universally and spontaneously adopted by the entire scientific community, would be a step forward for computational reproducibility.

However, none of these approaches for reproducible research works across all major formats, technologies and languages used in research. Current tools for reproducible research are tied to specific programming languages or platforms (e.g. iPython/Jupyter for Python, Sweave for R). Distributing results using a single format (e.g. as a virtual machine, a container solution like Docker or cloud infrastructure) limits the audience to only researchers who are able to access that platform; for instance, distributing results as Docker containers means that only researchers with the knowledge and equipment necessary to use the Docker container software are able to reproduce the work. Some researchers make their results available using cloud computing environments, but these are not equally available to all researchers globally due to limitations in national or organizational cloud infrastructure, or Internet bandwidth. These format access restrictions are a barrier to reproducibility, and hinder the sharing of computational results between researchers; this is the technological challenge we overcome in this work.

Furthermore, technological solutions to the problems of computational research reproducibility require a critical mass of researchers to be able to use them; the more widely a reproducibility tool is adopted, the greater the value in adopting it. In 'Building Towards a Future Where Reproducible, Open Science is the Norm'[17] Ram & Marwick summarise this:

> *[Until recently, researchers who put time and effort into documenting and sharing data and details of their analysis were considered outliers… Once a critical mass of scientists share their data and code, it would serve as a multiplier effect and allow disparate groups of researchers to rapidly solve problems]*

We assert that the lack of general tools for computational research reproducibility is the single biggest obstacle to its promotion across the research community. We further assert that the diversity of languages and technologies across the research community means that no specialized tool or approach will achieve this critical mass of general adoption; the consequence of this is that reproducibility is confined to within disciplinary boundaries. We believe that the research community can and should improve on this situation, and that the solution lies in common-denominator technical solutions using the most general of open-source technologies.

Here, we present an approach and examples for research reproducibility in computational research using a 'reference environment', a self-contained set of the minimal software needed to reproduce a published result. Using reference environments, computational research can be reproduced across multiple formats – as virtual machines, as containers and in cloud infrastructure – producing the same results in each format. The reference environment approach works for all technologies and languages and promotes distribution, evaluation and archiving of research results across the boundaries of discipline and specific technological expertise. Publication of a reference environment satisfies a number of the reproducibility criteria specified in the Institute for Computational and Experimental Research in Mathematics (ICERM) workshop report[18], specifically those concerning availability of code, parameters, and instructions for repeating computational experiments.

## Reference environments as a general solution to computational reproducibility across disciplines

The key value of the reference environment approach is that all reproducible research formats described in the Introduction can be easily produced from a single set of scripts, and the environments produced are functionally equivalent. This means that reviewers and readers can choose the format most appropriate to their situation and experience, while researchers publishing computational work can make it available to the broadest possible audience with minimal additional effort.

To *use* a reference environment, researchers can choose from a range of formats to download or access the environment, start it up, and execute code to reproduce a result (described in the Methods section under 'Using Reference Environments').

To *create* a reference environment, researchers choose a template environment to start with (e.g. with Python/R preinstalled), write a single set of installation scripts by filling in blanks in the basic scripts provided, add instructions to pull code from a public repository, and then choose deployment as any or all of a virtual machine, a Docker container, a cloud image and a read-only ISO image (described in the Methods section under 'Creating Reference Environments'). The most common reference environments are a set of code and open-source software packages built on top of the template environments we provide, which use a lightweight Linux operating system from a common family of distributions (Ubuntu); we chose this because of the large installed user base for this distribution, because of its (largely) free-software nature, and because it is the closest thing to a 'common' operating system in the research community; we believe that more results are more readily reproduced in this environment than in any other single operating system.

We also provide detailed step-by-step instructions on using and making reference environments at the companion website to this work: (http://uomsystemsbiology.github.io/research/reference-environments/).

To demonstrate the utility and broad applicability of the method, we present examples for major languages and technologies across the computational life sciences (Supplementary Table 6), including examples using R, MATLAB/Octave, Python, Java and C++. The code for each example is freely available, and readers are encouraged to experiment with these and use them as a base to build their own reference environments.

To show how reference environments can benefit discussions of reproducibility, we also include examples reproducing two network analyses for network link prediction[19] (Example 3) and network deconvolution[20] (Example 4). Both analyses were not reproducible at the time of their initial publication[5], but distributing a reference environment at the time of publication could have answered the specific questions raised post-publication, and provided a common objective reference point between authors, editorial staff, reviewers and readers. Finally, we show results from another study[21] to demonstrate that different versions of the common programming language MATLAB (as used by the network analysis publications) can produce different output values from the same data and code (Example 10). Since reviewers and readers may not execute code using exactly the same versions of software and packages, this could be interpreted as inability to reproduce published work, further emphasising the need for the detailed objective specification produced by a reference environment.

**Overview of constructing reference environments**

Figure 1 shows the workflow of constructing reference environments for a result. The first step is the construction of 'base environments' for each platform from a standard Linux distribution (A). Base environments include an operating system (for all but container platforms), and a minimal set of software components required to bootstrap the rest of the workflow. Our base environments and the scripts used to construct them are freely available and under version control, and researchers may choose to modify this to produce their own base environments, or create their own from scratch.

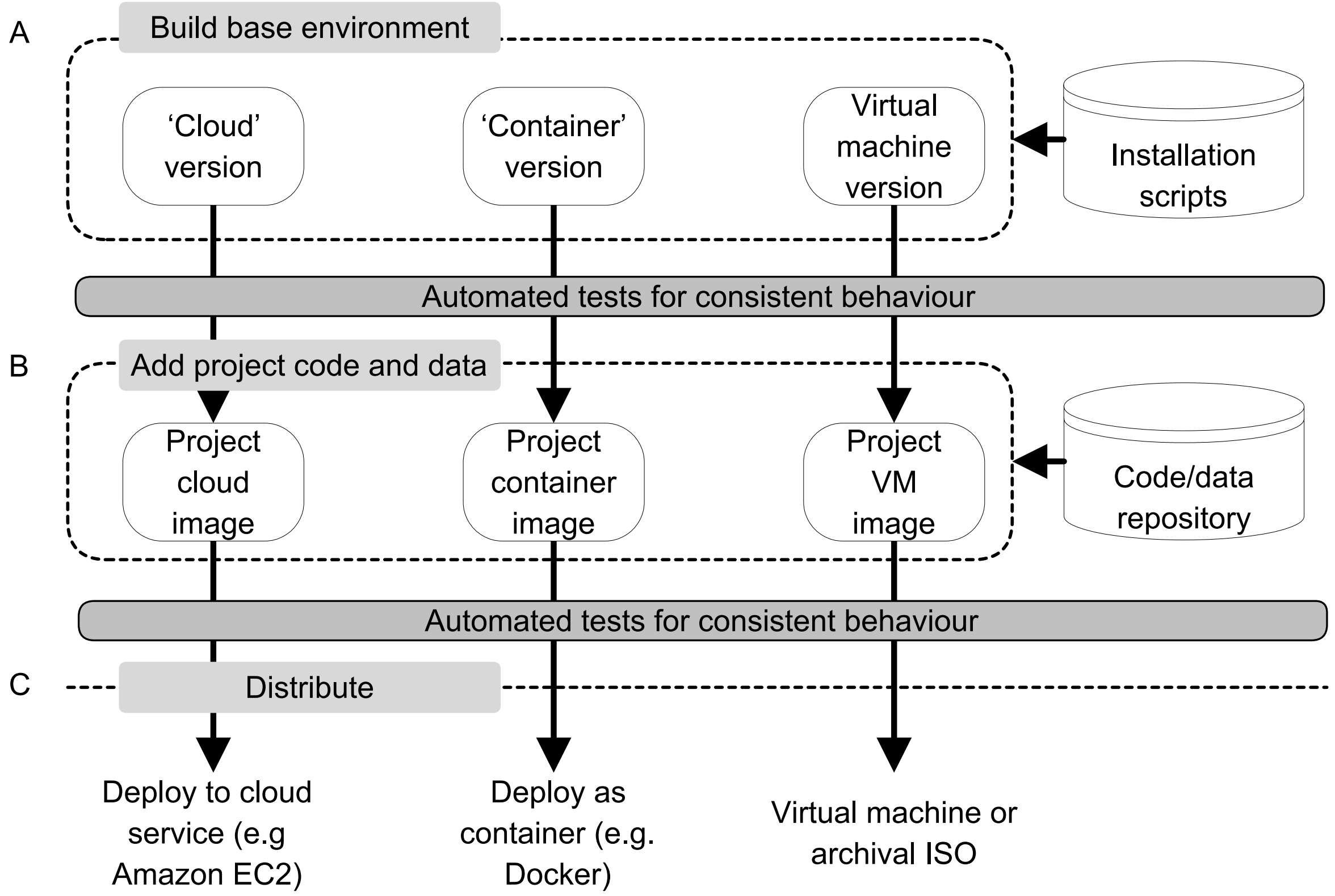

*Figure 1: Construction of reference environments. (A) Creation of a base environment using standard installation scripts. (B) Addition of project-specific code and data, followed by configuration and setup. (C) Distribution of the environment in equivalent formats across different platforms.*

Although many results can be replicated directly from the Linux command line, we have included a lightweight desktop, Lubuntu, (www.lubuntu.net) to maximize accessibility and ease of use for researchers who do not use command-line tools. This is a pragmatic policy decision which we encourage other reference environment users to adopt; many researchers do not use command-line tools on a regular basis, and wide and active engagement is critical for reproducibility techniques to achieve common adoption.

Next, the code and data specific to a result are added to each platform (B). This step may include compiling project code or performing platform-specific configuration (e.g. setting up a desktop environment for results that require interaction through a user interface). Templates for these scripts are provided, along with all scripts for all examples shown.

Finally, completed reference environments can be released through direct download for virtual machines, or from a central repository for cloud-based or container systems (C). To support scholarship and accurate referencing, DOIs can be produced for the code generating the environments and for the digital artifacts themselves (see examples in Supplementary Table 6).

**Self-contained environments vs. provisioned environments**

Reference environments come in two types: **self-contained** environments which include all the data and code required to reproduce a result, and **provisioned** environments which need to download data and code from the Internet the first time they are started. The source of the reference environment will usually tell you whether it is a self-contained or provisioned environment.

- **Self-contained environments** generally come in the format of a bootable ISO image Docker container. Once you have downloaded them, they do not depend on access to the Internet to work, or on the future availability of data or code resources, meaning that they will generally produce exactly the same results in future as they do now. Also, they are usually not ‘permanent’, so changes made to the environments will not be retained. Self-contained environments are ideal for *replicating* or *recomputing* a result.
- **Provisioned environments** generally come in the format of a virtual machine (VirtualBox VM) or cloud (Amazon EC2) instance. Each of these formats builds the reference environment ‘on-demand’: they start with a ‘base box’, then install and configure the project software and data. This means that to work they require access to the Internet, and for project-specific resources to be available (for instance, tools and packages must be accessible via online repositories). It also means that if resources change (e.g. a new version of a software tool is released), they may not produce exactly the same results in future as they do now. Once set up they are permanent environments in which changes are retained, and the process of installing and configuring project software and data should be similar to the process a researcher goes through when configuring their physical machine. Provisioned environments are ideal for *analyzing* or working with a result. Having a worked example of software installed and running on a machine can help with troubleshooting installation problems, and with testing the effect of configuration or software changes.

**Applications of reference environments**

Reference environments can be applied to solve a number of reproducibility problems in computational biology research and we describe and present two applications below. Further examples for each of these applications are detailed in the Supplementary Information. Each example is available as a virtual machine, a cloud-based Amazon EC2 instance, and as an archival ISO image. Examples which can be run from the command-line only and do not require a GUI are also available as Docker containers.

First, <u>distributing</u> computational results, as a simple and reliable way to deliver all the software and configuration associated with a particular result. A reference environment that generates the major conclusions and figures associated with a publication allows researchers to quickly run the relevant software and investigate the results in detail, reducing the time spent on installing and configuring software.

Second, <u>archiving</u> computational results for replication in perpetuity. Results that rely on tools and resources available over the Internet can change if those tools and resources change; they may become unavailable in future, or give different output. To make sure a result produced at one point in time continues to be reproducible, a read-only ISO image (ISO9660 standard) of the complete reference environment can be created. This image is a self-contained bootable operating system, which does not rely on access to the Internet to reproduce a result, and can be run in a virtual machine or off a USB flash drive. Images demonstrating this approach have been created for all the reference environment examples described in Supplementary Table 6.

## Methods

### Using reference environments

Reference environments are a complete software stack which reproduces a specific result or set of results; their overall goal is to make the process of *replicating and exploring computational biology research* easier and more reliable. Reference environments are available across four different formats:

- Virtual machines using VirtualBox (www.virtualbox.org)
- Amazon EC2 cloud images (https://aws.amazon.com/ec2)
- Docker containers (https://www.docker.com)
- Read-only bootable ISO images

Reference environments are built for each of these formats from a single set of installation scripts using the open-source tools Packer (http://packer.io/) and Vagrant (https://www.vagrantup.com).

A detailed set of step-by-step instructions for using reference environments in each format is provided in the Supplementary Information to this article, and online at the 'Guide to Reference Environments' (http://uomsystemsbiology.github.io/research/reference-environments/)

To learn how to use existing reference environments, refer to 'Using reference environments' in the Supplementary Information. A detailed list of example reference environments across different languages and technologies can be found in Supplementary Table 6.

### Reference environment formats

Using a reference environment typically involves running a single command or set of commands to execute code on the environment, generate some results and write those results to a local filesystem. Such results may be figures or tables from a publication, or they may be a more general demonstration for a program or piece of code. Because reference environments are produced in different formats, users can choose the format best suited to their specific situation and goals.

The **Bootable ISO** format is supported by most virtualization software across platforms and operating systems. It is read-only, meaning that changes are erased when the environment is rebooted, making it primarily suitable for verifying results or replicating the results of a previous computation.

**Docker containers** are also read-only, unless the container's state is saved. They are typically available from the Docker Hub, and are also primarily suitable for verifying results or replicating an existing analysis.

The **VirtualBox VM** format allows changes to persist until the environment is destroyed or deleted, and as a result is primarily useful for exploring the detail of a published result; for instance, making parameter or configuration changes and discovering how they affect the result, or testing the impact of code changes. This format is also useful for troubleshooting installation or configuration of software, as a 'known' good configuration.

The **Cloud (Amazon EC2)** instance format is also persistent, and is also useful for exploring the detail of a published result, especially in situations where the result requires computational resources not available in a local environment (for instance, high memory or disk space requirements).

More specific details on the differences between formats are given in Supplementary Table 1.

### Examples of reference environments

Reference environments have been implemented for a wide range of technologies and languages. Supplementary Table 6 lists a series of example reference environments for different projects and publications, and provides access information for the different formats. References are provided for publications where the reference environment was produced at the time of publication and submitted along with the manuscript.

### Creating reference environments

Reference environments are constructed in stages. Most reference environments start from a common base: a 32- or 64-bit Linux system built with Lubuntu (http://lubuntu.net), a lightweight variant of Ubuntu Linux.

Some reference environments also have the Lubuntu desktop installed, along with basic tools for viewing and manipulating results (text editor, picture viewer, web browser). The common base for each reference environment is built using a single set of scripts under version control; then project-specific code and data is added, and the environments are made available on the Internet. Figure 1 earlier summarizes the process of constructing and distributing reference environments.

The *core environment layer (A)* uses Packer to create equivalent base environments ('base boxes' in Packer terminology) across three platforms: VirtualBox VMs managed by Vagrant, Docker containers and Amazon EC2 instances. Bootable ISO-format reference environments are created later from the VirtualBox VM format.

The *project layer (B)* builds on top of these base boxes using Vagrant to add code and data for each specific research output. There is one Vagrant project for each research output and set of environments. Each Vagrant project contains a 'Vagrantfile', a plaintext configuration file which references the base box to use for the project, and calls a set of scripts to provision (build) the code and data for the research output.

In the *distribution layer (C)*, each version of the reference environment is pushed to the services which host it. VirtualBox VM and AWS cloud versions of an environment are available through Vagrant, and Docker containers are on the Docker Hub.

The process of creating a reference environment to reproduce analysis is summarized in Figure 2 and described in detail in the Supplementary Information, as well as in the Guide to Reference Environments online (http://uomsystemsbiology.github.io/research/reference-environments/) . Supplementary Table 7 also provides a checklist of key activities for creating a reference environment, for ease of adoption.

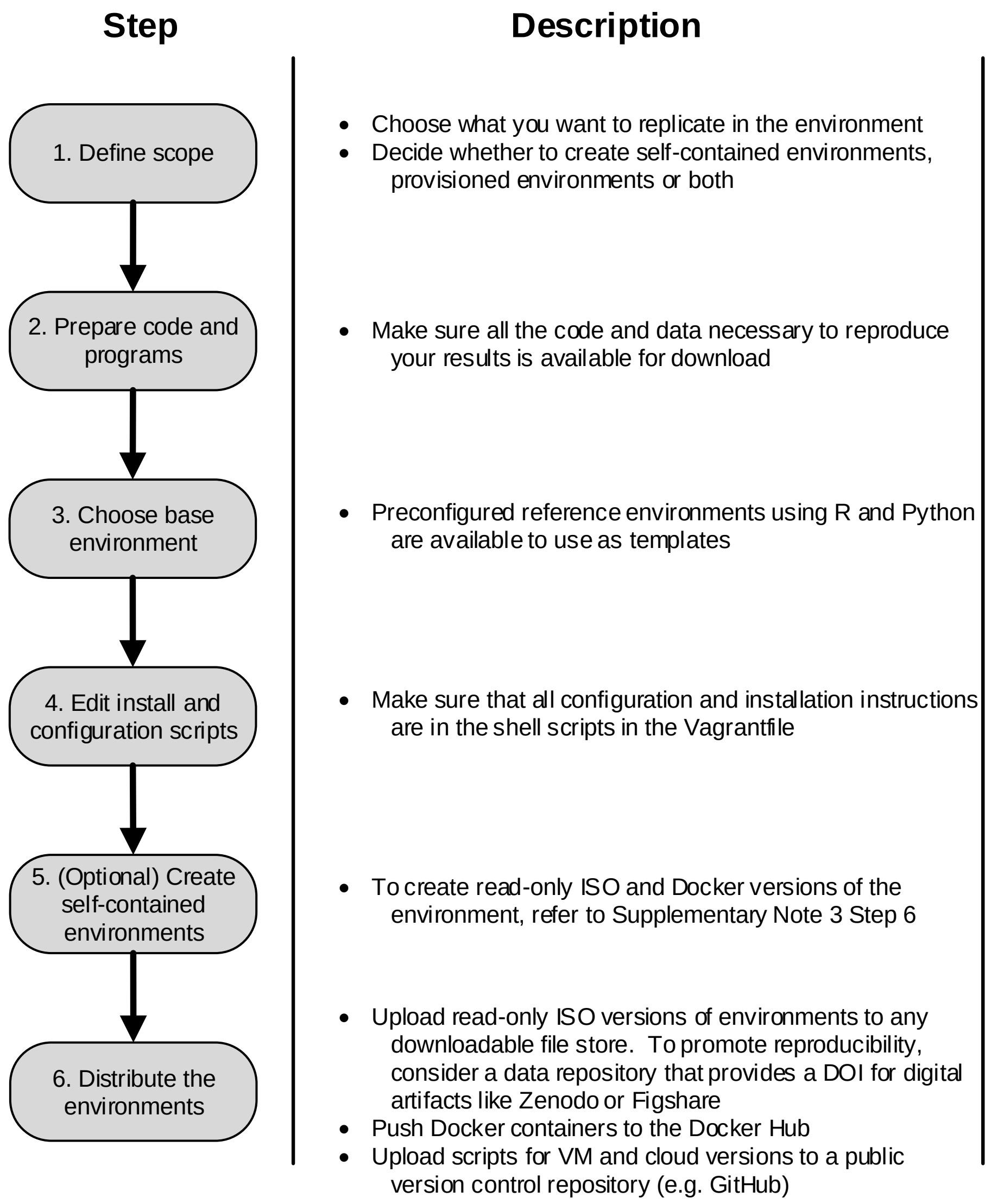

*Figure 2: Flowchart for creating a reference environment*

## Discussion

### Overcoming challenges to computational reproducibility using reference environments

When creating a reference environment to reproduce results, several factors must be assessed to ensure reproducibility. Some results are easier to reproduce than others, and in some cases a complete set of results cannot be reproduced in a reference environment because of restrictions of data size, computational complexity, data availability or licensing considerations.

Table 1 lists key considerations for defining the scope of reproducibility within a reference environment, and suggests mitigation strategies for common challenges to reproducibility. Partial reproducibility for a subset of results, or the final stages of a computation, still has considerable value if it represents an example of the software method correctly set up and configured. In addition, we recommend that computational publications include clear explanations of which results can be easily replicated in a reference environment, and which cannot, and for what reasons.

| Issue | Response |
|---|---|
| **Computational load:** Result requires high-performance computing resources to reproduce | Consider limiting scope to post-processing after the step requiring high-performance computing resource. Make intermediate datasets publically available, and design the reference environment to download and process them. |
| **Architecture requirements:** Result requires specialized architecture (e.g. parallel processing, low-latency system) to reproduce | Consider simulating the architecture and reproducing a reduced version of the result, or an example result derived from test data, to illustrate the dependence on architecture. |
| **Data size:** Core data for a result are too large for users to reasonably download | Reproduce a partial result from a subset of the data, or implement batch processing so that results can be delivered in sections. Also consider limiting scope to post-processing if intermediate datasets are small enough to download. |
| **License restrictions for software:** Result requires commercially-licensed software to reproduce | Identify open-source alternatives for commercial software, or limit scope to parts of a result which can be reproduced without license restrictions. Consider negotiation with license holders: some companies may be willing to produce limited-use licenses for the purpose of reproduction. |
| **Data availability:** Data are confidential, embargoed or restricted in access | Consider producing an example result derived from test or publically-available data to illustrate the core method. Explore options for anonymizing data if appropriate to allow public distribution. |

*Table 1: Issues affecting scope definition for a reference environment*

### Integration of reference environments with other tools, platforms and standards associated with reproducible research

Many tools and platforms relate to the general concept of ‘reproducible’, ‘recomputable’, or ‘replicable’ research. Table 2 lists some of the most common tools, platforms and standards relevant to reproducible research, and describes how reference environments can integrate with or make use of them.

| Type | Examples | Relationship with reference environments |
|---|---|---|
| **Workflow systems** which record, manage and execute a series of steps in a computation | Taverna<br>Galaxy<br>caWorkbench<br>Vistrails<br>Kepler<br>Pegasus<br>Sumatra | Run workflow systems within a reference environment to demonstrate a specific workflow, or as a worked example of installation and configuration. |
| **Literate programming** approaches, which integrate executable code with human-readable documents | iPython/Jupyter (Python)<br>knitr/Sweave (R)<br>LaTeK | Reference environments can recreate analyses and papers using literate programming tools. |
| **Software containers**, which separate software processes on a single machine for security, stability and resource management purposes | Docker<br>Rocket | Docker images are produced as part of the reference environment process; other container tools can be supported in future. |
| **Standards and formats** for reporting and encoding models and experimental details | MIAME<br>MIABI<br>MIAPE<br>MIASE<br>MIRIAM<br>CellML<br>SBML<br>SBGN<br>SED-ML | Research outputs using reference environments can adhere to specific standards and formats for processing or generating results. |
| **Repositories** for storing method and protocol information, data and environments | protocols.io<br>OSF.io<br>Runmycode.org<br>Dryad<br>nucleotid.es<br>Bioboxes<br>GigaDB<br>GigaGalaxy<br>Figshare<br>Zenodo<br>Dataverse<br>JOVE<br>MyExperiment | Repositories allowing automated access to information and data can be used as sources for provisioning reference environments. ISO versions of reference environments can be made into citable research outputs using DOIs. |

*Table 2: Relationships between reference environments and other reproducible research tools*

**Conclusion**

Reproducibility in the computational sciences presents challenges of science culture, and of community and infrastructure support, as well as technological challenges. The method we describe for creating reference environments provides a comprehensive and general solution to common technological problems of reproducibility in computational biology research, overcoming the barrier to adoption of reproducibility approaches based on a single technology or language.

It explicitly separates the core scientific findings of a piece of computational research from the software implementation in which they are embedded, and allows readers and reviewers to choose the most appropriate format for their situation. Reference environments integrate elegantly with other reproducibility approaches, and also have many potential applications in more general reproducibility; providing standard environments for testing or benchmarking, or in teaching to accelerate students' hands-on experience with tools and languages.

# Supplementary Information

## 1. Reference Environment Formats

Four different formats for reference environments are summarized in Supplementary Table 1, along with requirements, permanence for the format, suggested application, and source. Below, use of each of these Reference Environment formats is discussed in detail.

*Supplementary Table 1: Summary of formats for reference environments*

| Format | Requirements | Permanence | Application | Source |
|---|---|---|---|---|
| **Bootable ISO** | Most virtualization software | Read-only - changes are erased on reboot | Replication/ recomputation<br>Verifying results | Direct download |
| **Docker container** | Docker v1.8.1 or greater | Read-only unless container is committed | | Docker Hub |
| **VirtualBox VM** | VirtualBox v4.3.10 or greater<br>Vagrant v1.7.2 or greater<br>Project-specific resources on startup | Changes persist until the environment is destroyed/deleted | Exploring results<br>Testing the impact of changes<br>Troubleshooting installation | Hashicorp Atlas (via Vagrant) |
| **Amazon EC2 instance** | Vagrant v1.7.2 or greater<br>Amazon Web Services EC2 account<br>Project-specific resources on startup | | | Amazon EC2 (via Vagrant) |

## 2. Using Reference Environment Formats

This section provides step-by-step instructions for using each of the reference environment formats described in Supplementary Table 1.

### 2.1 Using a bootable ISO reference environment

To use a bootable ISO reference environment, users need virtualization software installed. Almost all virtualization tools can boot from an ISO image, but common examples are Oracle VirtualBox or VMWare (http://www.vmware.com) for Windows, OSX and Linux, or QEMU (http://wiki.qemu.org/Main_Page) for Linux. Supplementary Table 2 describes the steps to use a bootable ISO format reference environment with VirtualBox v4.3.10 or later.

*Supplementary Table 2: Using a reference environment in bootable ISO format*

| Step | Description | Action |
|---|---|---|
| **1** | Download the ISO file | ISO images may be available on a data repository like Figshare (http://figshare.com) or Zenodo (https://zenodo.org), or they may be available through a journal website. |
| **2** | Create a virtual machine to boot the ISO | Create a new virtual machine by clicking the 'New' button, and for the operating system choose 'Linux Ubuntu' 64- or 32- bit, depending on the ISO. |
| | | Set the memory size to 2048Mb; this should be enough for most reference environments, although some may specify more.<br>Choose not to add a virtual hard drive, since the machine will boot off the ISO image. |
| **3** | Mount the project ISO | Select the new virtual machine and click Machine\|Settings on the menu. In the Settings window, click 'Storage'. Under 'Attributes', click the CD-ROM icon and select 'Choose a virtual CD/DVD file'. Find the project ISO file and click 'Open'. Click OK to close the Settings window. |
| **4** | Start the machine | Click the Start button on the menu bar. When the boot menu appears, choose 'Boot the live system' from the boot menu. The machine will start up, automatically log in and show the desktop. |
| **5** | Execute the project instructions | Most reference environments will have instructions directly on the desktop which tell you how to execute the project code; this will normally involve running a script on the desktop by double-clicking it. The script should run and a message should appear describing where the project outputs have been written. |

### 2.2 Using a Docker container reference environment

To use a Docker container reference environment, Docker v1.8.1 or greater must be installed. Supplementary Table 3 describes how to use a reference environment in the Docker container format.

*Supplementary Table 3: Using a reference environment in Docker container format*

| Step | Description | Action |
|---|---|---|
| **1** | Get the Docker image for the project | If the image is on the Docker Hub, run the command:<br>`sudo docker pull uomsystemsbiology/<projectname>`<br>The image will download and be ready to run. |
| **2** | Execute the project instructions | Most projects will have a shell script suitable for use with the `docker run` command: consult the project documentation for the specific syntax. |

### 2.3 Using a VirtualBox VM reference environment

To use a VirtualBox VM reference environment, users need VirtualBox v4.3.10 or greater and Vagrant v1.7.2 or greater installed. Supplementary Table 4 describes the steps to use a reference environment in VirtualBox VM format.

*Supplementary Table 4: Using a reference environment in VirtualBox VM format*

| Step | Description | Action |
|---|---|---|
| **1** | Get the Vagrant code for the project | If git is installed, clone the project from GitHub:<br>`git clone https://github.com/uomsystemsbiology/<project-name>.git` |
| | | Otherwise, go to the project on GitHub, using the link: https://github.com/uomsystemsbiology/<project-name> and choose 'Download Zip' from the project repository page. |
| **2** | Go to the Vagrant-managed project directory | If the repository was cloned, change to the <project-name> directory. If the zipped archive was downloaded, extract it and change to the created directory in command prompt. |
| **3** | Bring up the Vagrant managed environment | Type `vagrant up` at the command prompt. The virtual machine will start and provisioning scripts will run. |
| **4** | Access the environment | The environment can be accessed using the desktop if one is provided, or using the command `vagrant ssh`. |
| **5** | Execute the project instructions | If a desktop is provided, most reference environments will have instructions directly on the desktop; this will normally involve running a script on the desktop by double-clicking it. The script should run and a message should appear describing where the project outputs have been written.<br>If the machine is accessed using `vagrant ssh`, consult the project documentation for the specific syntax. |

### 2.4 Using an Amazon EC2 cloud reference environment

To use an Amazon EC2 reference environment, users need Vagrant v1.7.2 or greater installed, and an Amazon Web Services EC2 account. AWS offers a free tier which should be enough to run most reference environments, but using more than the free tier's allocated storage space incurs charges, so users should be careful to delete the environments once they are no longer needed. Supplementary Table 5 describes the steps to use a reference environment in Amazon EC2 format.

*Supplementary Table 5: Using a reference environment in Amazon EC2 instance format*

| **Step** | **Description** | **Action** |
|---|---|---|
| **1** | Get the Vagrant code for the project | If git is installed, clone the project from GitHub::<br>`git clone https://github.com/uomsystemsbiology/<project-name>.git`<br>Otherwise, go to the project on GitHub, using the link:<br>https://github.com/uomsystemsbiology/<project-name><br>and choose 'Download Zip' from the project repository page. |
| **2** | Go to the Vagrant-managed project directory | If the repository was cloned, change to the <project-name> directory. If the zipped archive was downloaded, extract it and change to the created directory in command prompt. |
| **3** | Create environment variables for the AWS access keys. | The keys are called AWS_ACCESS_KEY_ID, AWS_SECRET_ACCESS_KEY, and AWS_SECURITY_GROUP. Keys are received when an AWS account is created, and a security group needs to be created before any instances are launched.<br>Creating environment variables varies by operating system, but generally in Linux and OSX they are added by modifying `/etc/environment` or `/etc/profile`. In Windows |
| | | they are added using the desktop through Control Panel/System/Advanced System Settings. |
| **4** | Bring up the Vagrant managed environment | Type `vagrant up --provider=aws` at the command prompt. The virtual machine will start and provisioning scripts will run. |
| **5** | Access the environment | The environment can be accessed using the command `vagrant ssh`, or `vagrant rdp` if a Remote Desktop Protocol client is installed . |
| **6** | Execute the project instructions | If a desktop is provided, most reference environments will have instructions directly on the desktop; this will normally involve running a script on the desktop by double-clicking it. The script should run and a message should appear describing where the project outputs have been written.<br><br>If the machine is accessed using `vagrant ssh`, consult the project documentation for the specific syntax. |

## 3. Examples of reference environments

This section provides links to examples of publically available reference environments replicating published work, across the most common languages and platforms in the computational life sciences.

*Supplementary Table 6: Citation and access details for example reference environments*

| # | Example | Languages used |
|---|---|---|
| 1 | Machine learning approaches to modelling eukaryotic transcription[1]<br><br>**Bootable ISO image:** https://dx.doi.org/10.5281/zenodo.30377<br><br>**Docker container:** https://hub.docker.com/r/uomsystemsbiology/budden2014predictive/<br><br>**Vagrant VM:** https://github.com/uomsystemsbiology/budden2014predictive_reference_environment | R |
| 2 | Hormonal regulation of renal excretion[2]<br><br>**Bootable ISO image:** https://dx.doi.org/10.5281/zenodo.30035<br><br>**Vagrant VM:** https://github.com/uomsystemsbiology/hormonal_regulation_of_renal_excretion | R, OCaml |
| 3 | Network link prediction<br><br>**Bootable ISO image:** https://dx.doi.org/10.4225/49/55DA8FA8CE707<br><br>**Docker container:** https://hub.docker.com/r/uomsystemsbiology/barzel2013network/<br><br>**Vagrant VM:** https://github.com/uomsystemsbiology/barzel2013network_reference_environment | MATLAB |
| 4 | Network deconvolution<br><br>**Bootable ISO image:** http://dx.doi.org/10.5281/zenodo.29668<br><br>**Docker container:** https://hub.docker.com/r/uomsystemsbiology/feizi2013network/<br><br>**Vagrant VM:** https://github.com/uomsystemsbiology/feizi2013network_reference_environment | MATLAB |
| **5** | Cardiac cell calcium release modelling[3]<br>**Bootable ISO image:** http://dx.doi.org/10.5281/zenodo.32916<br>**Vagrant VM:** https://github.com/uomsystemsbiology/ryr_simulator_reference_environment | MATLAB, R |
| **6** | Bond graph modelling of biochemical networks[4]<br>**Bootable ISO image:** https://dx.doi.org/10.5281/zenodo.29623<br>**Docker container:** https://hub.docker.com/r/uomsystemsbiology/hbgm/<br>**Vagrant VM:** https://github.com/uomsystemsbiology/hbgm_reference_environment | Octave, LaTeK |
| **7** | ERK-MAPK signal transduction modelling in human epidermis[5]<br>**Bootable ISO image:** https://dx.doi.org/10.5281/zenodo.33361<br>**Docker container:** https://hub.docker.com/r/uomsystemsbiology/epidermal_data/<br>**Vagrant VM:** https://github.com/uomsystemsbiology/epidermal_data_reference_environment | Python, MATLAB |
| **8** | Parallel data mining using WEKA[6]<br>**Bootable ISO image:** https://dx.doi.org/10.5281/zenodo.22415<br>**Docker container:** https://hub.docker.com/r/uomsystemsbiology/flexdm/<br>**Vagrant VM:** https://github.com/uomsystemsbiology/flexdm_vagrant | Java |
| **9** | Cell modelling using CHASTE (‘Cancer, Heart and Soft Tissue Environment)<br>**Bootable ISO image:** http://dx.doi.org/10.5281/zenodo.33077<br>**Vagrant VM:** https://github.com/uomsystemsbiology/chaste_vagrant | C++, Python |
| **10** | Comparison of MATLAB code using different versions of MATLAB<br>Bootable ISO image: https://dx.doi.org/10.5281/zenodo.31215<br>Vagrant VM: https://github.com/uomsystemsbiology/version_comparison_reference_environment | MATLAB |

## 4. How to create reference environments

This section provides step-by-step instructions on how to create a reference environment to replicate computational work, following the sections described in Figure 2 in the main text.

### 4.1 Define the scope of the environment

To decide how to replicate an analysis or result in a reference environment, think about which parts of the analysis provide the strongest supporting evidence to the claims you are making, and what kind of output presents them in the most convincing manner. Reference environments do any and all of the following:

- Output individual values on the command line showing the result of a single computation
- Output text files to the filesystem
- Output plots in PDF or bitmap form and open them on the desktop
- Compile and run programs which open windows on the desktop

Which of these is the best way to demonstrate a result depends on your specific situation. Good reference environments typically produce output which is clear and compelling, like a figure in a publication or talk, but is also backed up with detail, so that readers and reviewers can investigate further if they want to. To do this, consider generating visual output (plots) together with data output (text files), and tell users that this has been done.

You also need to decide how much of an analysis will be replicated in the reference environment. In an ideal situation, every single part of an analysis starting from some form of 'raw data' would be replicated in a reference environment, but for many results this may not be realistic or desirable – it may take too long to run, require large amounts of one resource type or another, or not be possible because of licensing or privacy restrictions. Table 1 in the main text lists the most common issues to consider when defining scope for a reference environment and suggests responses for each one.

Finally, consider whether you want to create *self-contained* versions of your reference environment as well as *provisioned* versions. As described in the main text, **self-contained** environments include all data and code required to reproduce a result, while **provisioned** environments need to download data and code from the Internet the first time they are started.

- **Self-contained environments** generally come in the bootable ISO image or Docker formats. Once you have downloaded them, they do not depend on access to the Internet to work, or on the future availability of data or code resources, meaning that they will generally produce exactly the same results in future as they do now. Changes made to the environments will not be retained after 'restarting' the virtual machine but self-contained environments are ideal for *replicating* or *recomputing* a result.
- **Provisioned environments** generally come in the virtual machine (VirtualBox VM) or cloud (Amazon EC2) formats. Each of these formats builds the reference environment 'on-demand': they start with a 'base box', then install and configure the project software and data. This means that they require access to the Internet, and for project-specific resources to be available (for instance, tools and packages must be accessible via online repositories). It also means that if resources change, they may not produce exactly the same results in future as they do now. Once set up, they are permanent environments in which changes are retained, and the process of installing and configuring project software and data on them should be similar to a researcher's physical machine. Provisioned environments are ideal for *analyzing* or working with a result. Having a worked example of the software installed on a machine can help troubleshoot installation problems, and with testing the effect of configuration or software changes.

Self-contained environments have the advantage that they are a permanent record of your result, but they also require that all the code and data required to produce a result be present in the environment, and this may not be possible or desirable for the reasons described in Table 1. When you create self-contained or provisioned environments, we recommend that you clearly explain the difference in the documentation for a reference environment, and in the accompanying publication.

### 4.2 Prepare your code and programs for being run in a reference environment

To make your analysis suitable to run in a reference environment, you will need to make the entire installation and analysis process scripted and non-interactive. To do this, work through the following steps:

- Make a list of all the *dependencies* it requires – all the programs, tools, packages and data needed to run it.
- Define and list all the commands required to install these dependencies on a basic Linux-based system in a non-interactive fashion (e.g. using apt-get commands). Users will not be able to interact with the system while it is provisioning, so this process needs to be completely non-interactive: look for 'quiet' or 'hands-off' options in each step of the installation process.
- Structure your code or analysis so that it can be executed non-interactively. This might mean rewriting particular parts, or it might mean creating a 'run_experiments.sh' script which runs a series of analyses from the command-line one after another.
- Make sure the code and data (if required) are available at a source from which you can script the installation. If you make this available at a public code or data repository, this means you can make a provisioned environment as described above – when the provisioned environment starts up, it will get the code or data from the public repository. This is the most accurate replica of what a user would do if they were trying to replicate your analysis on their own machine.
- If you do not make your code available at a public repository, you can still create self-contained environments with the code/data in it, but you will not be able to make provisioned environments without providing special access (e.g. by creating a separate provisioning user with limited access to your private repository).

### 4.3 Choose a base environment from which to start

The easiest way to develop a reference environment is to start from one of the preconfigured template environments. There are three available:

**Blank reference environment template:**
https://github.com/uomsystemsbiology/reference_environment_template

**R reference environment template:**
**https://github.com/uomsystemsbiology/r_reference_environment_template**

**Comes with R preinstalled**

**Python reference environment template:**
**https://github.com/uomsystemsbiology/python_reference_environment_template**

**Comes with Python preinstalled**

Each of these templates includes sample scripts and suggested text for each part of the environment.

**Note that the default user 'sbl' and password 'sbl' has administrator access to all environments produced from the reference environment templates. These environments are therefore not secure, and should not be used to store confidential data. If a secure reference environment is required, edit the installation scripts to remove this user after installation.**

If none of these environments are suitable you can also develop your own 'base box' using Packer. Developing base boxes requires Packer v0.7.5 or greater installed. To make VirtualBox and Amazon EC2 format base boxes available, a Hashicorp Atlas (https://atlas.hashicorp.com) account is also required to host the base box. To make Docker base images available, a Docker Hub account (https://hub.docker.com) is required.

Once the accounts above are created, clone the Packer templates for creating the Systems Biology Lab boxes from GitHub:

https://github.com/uomsystemsbiology/vre_base

https://github.com/uomsystemsbiology/vre_base64

Scripts for building the base boxes are in the /scripts directory. The default user for environments produced by the Systems Biology Lab is `sbl` - users can be changed or added by editing the 'setup_users.sh' script.

With a configured Docker account, the Packer post-processors will export the Docker image to the Docker Hub. With a configured Atlas account, they will export the AWS and VirtualBox base boxes to Atlas.

#### 4.4 Edit the install and configuration scripts to get your code, resources and data into the environment

- If you have a git client installed and you are using one of the template reference environments then you can use the clone command:

  ```
  git clone https://github.com/uomsystemsbiology/
  reference_environment_template.git project-name
  ```

- Otherwise, you can use the 'Clone or download' link on the Github page to download a ZIP archive for the environment template.
- If you have developed your own base boxes, then edit the Vagrantfile to replace the values of `override.vm.box` and `docker.image` with the names of the new base boxes.
- Edit `2_install_core.sh` to add the instructions you identified in Step 2 for installing specific packages or code needed for the environment. To list the packages installed by default in the base box, bring up the template reference environment using `vagrant up`, then `vagrant ssh` into the machine and use `dpkg -l` from the command line. `2_install_core.sh` also usually populates `data/build_info.txt` which includes build information for the specific version of the code.
- Edit `3_install_desktop.sh` to add instructions for anything specific to the desktop environment. Remember that the Docker version of a reference environment, if you make one, typically doesn't have a desktop, so graphical or interactive tools/results won't work.
- Edit `4_configure_core.sh` to set up applications and compile code.
- Edit the existing `data/run_experiments.sh` script or write your own; most reference environments include one or more shell scripts which execute the commands required to reproduce a particular result.
- Edit the text files in the `/data` directory to specify the project name and organisation name, and some instructions to be written to the desktop wallpaper.

#### 4.5 (Optional) Create self-contained ISO and Docker versions of the environment

- If you want to make a read-only ISO of the reference environment, edit `data/remastersys.conf` to change the name of the output ISO file (the CUSTOMISO parameter) and the name displayed when the CD is booted (LIVECDLABEL). To create the ISO, uncomment the lines:

  ```
  #if !(is_docker)

  #config.vm.provision "shell", path: "scripts/make_iso.sh",
  privileged: false

  #end
  ```

  in the project Vagrantfile. When the environment is launched with `vagrant up`, this script will create an ISO in the `output` directory.

- If you want to make a Docker image of the environment, make sure you have Docker v1.8.1 or greater installed, and launch the environment with `vagrant up –provider=docker`.
- When the environment has started, tag the image using `docker tag`, and push the image to the Docker hub using `docker push`. The Docker documentation here (https://docs.docker.com/engine/getstarted/step_six/) has detailed instructions on using Docker commands.

### 4.6 Distribute the environment

Once reference environment scripts are finished and the created environments have been tested, they can be made available in a range of different ways. Table 3 below lists suggested distribution methods for each type of output.

| Environment type | Recommended distribution method |
|---|---|
| **Vagrant-managed virtual machine, Amazon EC2 cloud instance** | Make the Vagrantfile and the provisioning scripts available on a public version control repository (e.g. GitHub, SourceForge, BitBucket). |
| **Docker image** | Make the image available on the Docker Hub using `docker tag`, and `docker push` |
| **Bootable ISO** | Upload to any public download service. For accessibility and reproducibility, consider a generalised data repository like Figshare (https://figshare.com/) or Zenodo (https://zenodo.org/) which provide a DOI for digital artifacts, making citation easy. |

*Table 3: Distributing reference environments in different formats*

## 5. Ensuring reproducibility using reference environments

Creating a reference environment usually involves installing software and packages from the Internet. To make sure that reference environments produce consistent results in future, follow these guidelines:

- All commands to install packages, resources or make changes to the environment should be in the installation scripts (in the `/scripts` directory), or in scripts called by these scripts. Avoid using other methods to install packages or resources, such as installation of packages when R/Python project code is run for the first time. Keeping all environment setup and installation centralized in one location allows users to see all the changes and dependencies required to reproduce a result. If changes are made in other places, users may not realise they are happening and are important for reproducing a result.
- When installing packages or code in a script, specify a version to be installed. For instance, when installing operating system packages in Lubuntu, use `apt-get install package=version` to install a specific numbered version of the package. When getting code from a version control repository like GitHub, specify a particular release of the code. If you specify the *latest* release of the code, the environment may change in future and produce different results when run at different times.
- For self-contained environments (generally ISO images and Docker containers), remind users that these environments are a snapshot, and that changes made in them will not be saved. You can put reminders on the download page for the environment, or edit the Vagrantfile to put different information in the environment for each environment type.
- For provisioned environments (generally VM and cloud images), remind users that these require access to the internet to set up, and be clear about the version of packages or code that they install.

If they install the latest version of packages or code, warn users that this may mean results differ from published ones obtained using a previous version of a resource.

## 6. Checklist of activities for creating a reference environment

*Supplementary Table 7: Checklist of activities for creating a reference environment*

| | |
|---|---|
| ☐ | All installation and configuration instructions for packages and resources are in shell scripts in the `/scripts` directory |
| ☐ | Instructions for downloading code from a public repository (e.g. GitHub) are in shell scripts in the `/scripts` directory |
| ☐ | Version and build information in `data/build_info.txt` is being correctly set from `scripts/2_install_core.sh`. |
| ☐ | Configuration for applications is set, and code is being compiled, if appropriate, from `scripts/4_configure_core.sh`. |
| ☐ | Scripts that users will run to execute code and generate results are in `data/run_experiments.sh` |
| ☐ | Text files in `/data` have been edited to specify the project and organisation name, and instructions for reproducing results |
| ☐ | (Optional) Username and password for the reference environment OS have been changed |
| ☐ | (Optional) Read-only ISO of the reference environment has been produced (Figure 2 Step 6) |
| ☐ | (Optional) Docker container version of the reference environment has been produced (Figure 2 Step 6) |
| ☐ | Environments have been distributed to downloadable repositories (Figure 2 Step 7) |